# A survival model for course-course interactions in a Massive Open Online Course platform

Edwin H. Wintermute*[1], Matthieu Cisel[1] & Ariel B. Lindner[1]

[1]Paris Descartes University, Center for Research and Interdisciplinarity (CRI), 8 Rue Charles V, 75004 Paris, France
*To whom correspondence should be addressed.

## ABSTRACT

**Massive Open Online Course (MOOC) platforms incorporate large course catalogs from which individual students may register multiple courses. We performed a network-based analysis of student achievement, considering how course-course interactions may positively or negatively affect student success. Our dataset included 378,000 users and 1,000,000 unique registration events in France Université Numérique (FUN), a national MOOC platform. We adapt reliability theory to model certificate completion rates with a Weibull survival function, following the intuition that students "survive" in a course for a certain time before stochastically dropping out. Course-course interactions are found to be well described by a single parameter for user engagement that can be estimated from a user's registration profile. User engagement, in turn, correlates with certificate rates in all courses regardless of specific content. The reliability approach is shown to capture several certificate rate patterns that are overlooked by conventional regression models. User engagement emerges as a natural metric for tracking student progress across demographics and over time.**

## SIGNIFICANCE STATEMENT

What happens to a user's chances of successfully completing an online course when they also co-register for other courses? To answer this question, we built a mathematical model and fit it to data from 378,000 online course users. Our model makes use of concepts from reliability theory normally used to model the failure of mechanical devices over time. The effect of course co-registration on certificate rates follows a simple pattern that depends mainly on a single parameter called user engagement. Some courses tend to have highly engaged users. High engagement users, in turn, show increased certificate rates for all courses they register. Engagement levels can be estimated from a user's registration profile, making them useful for tracking user progress.

## INTRODUCTION

In recent years millions of students have registered for thousands of newly created online courses with topics spanning the range of human knowledge(*1*). Massive Open Online Courses (MOOCs) are a subset of online courses defined by a commitment to open access and unlimited registration(*2*). MOOC use is increasingly mediated though MOOC platforms, websites that offer centralized access to many courses through a standard user interface. The rising popularity of MOOCs motivates the detailed study of user outcomes.

The large user-bases and digital format of MOOCs generates large data sets in which a variety of studies have sought the keys to user success. Previous work has identified course-specific features of style and content that characterize highly effective MOOCs(*3-6*). Other studies look outside the course for student-specific demographic and social factors that affect performance(*7-9*). Here we consider the interaction-specific factors that come into play when users register for multiple courses. Although the central organization of MOOC platforms encourages multiple registrations, there has not yet been a systematic study of course-course effects.

Our data is collected from France Université Numérique (FUN), the French national MOOC platform. In a 23-month period, the platform logged 1,000,000 registrations from 378,000 unique users to 140 courses. Each registration event records a user ID, time stamp, and whether a certificate was obtained. On average, 8.1% of course registrations produced a certificate.

The framework of this study is a statistical model for the probability that a given registration event will produce a certificate. Each event is assigned a course-dependent term we call difficulty and a user-dependent term called engagement. A user's engagement term is estimated using their complete registration profile and therefore reflects the influence of each registered course on every other.

Tracking difficulty and engagement separately, we follow the progress of users who return to FUN for multiple MOOCs. Returning users are shown to be both more engaged and more inclined to register difficult courses. Similarly, changes in engagement and difficulty combine to drive increases in certificate rates for older users.

Our model is structured as a Weibull survival function, a formal framework commonly used to describe failure rates in mechanical systems with many independent modes of failure. In this way, the certificate rates of MOOC users are connected to the well-developed statistical methods of reliability analysis. This approach is shown to outperform conventional logistic regression in describing key global patterns in certificate rate.

## RESULTS

### The *βEND* model parameterizes a Weibull survival function

Figure 1 reviews the structure of the *βEND* model. Each course is assigned a difficulty term, *D*, that summarizes all course-specific features contributing to the certificate rate. High difficulty courses may exhibit, for example, advanced subject matter, heavy work-loads, or an unappealing presentation style. Each user is associated with an engagement term, $E_U$. High engagement users may enjoy natural aptitude, prior preparation, or a willingness to invest time.

We further assume that a user's engagement is determined by the set of courses they have chosen to register (Fig. 1A). Intuitively, we expect users to seek out courses that match their intrinsic level of skill and motivation. We therefore apply a course engagement term, $E_C$, for each course and calculate $E_U$ as the sum of $E_C$ for all courses registered by a user during the data collection period.

With these estimates of user engagement and course difficulty, we next sought an appropriate mathematical framework to describe the dynamics of user persistence or withdrawal. Student drop-out behavior is recognized as a complex social phenomenon with many causes(*10*). A course might be too advanced, too demanding, or too dull. Personal, professional or social circumstances can change(*11*). The many factors that keep a student actively working toward a certificate can be thought of as links in a chain. A single point of failure is sufficient for drop-out to occur.

The weakest-link concept is formalized by the Weibull survival function. In any system with multiple essential components, the failure time of the system is set by the minimum failure time among the components. The extreme value theory provides that, under appropriate conditions, the distribution of minimum failure times will approach a Weibull. This is true regardless of the model chosen for the failure of the individual components. For this reason, the Weibull distribution is mechanistically appropriate to describe failure rates in many complex systems(*12*).

The terms $E_U$ and *D* are therefore used to parameterize the survival function of a Weibull distribution. *D* becomes the time-to-failure threshold: a user must persist beyond *D* to obtain a course certificate. $E_U$ serves as the scale parameter for the distribution, with larger values indicating a greater density of long-term survivors. We assign $\beta$ as a Weibull shape parameter. Values of $\beta$ less than 1, as we discover in this dataset, indicate a progressively decreasing rate of failure.

$$S = exp\left[-\left(\frac{D}{E_U}\right)^{\beta}\right] \quad (1)$$

Finally, we consider the fact that students who register for many courses within a short time period are forced to divide their efforts. *N* represents the number of courses registered during a single registration burst event, defined below. We make the simple assumption that probability density is divided evenly among multiple registrations. This leads to the expression $C = S/N$, where $C$ is the expected probability of obtaining a certificate. A graphical interpretation of the Weibull survival function is offered in figure 1B.

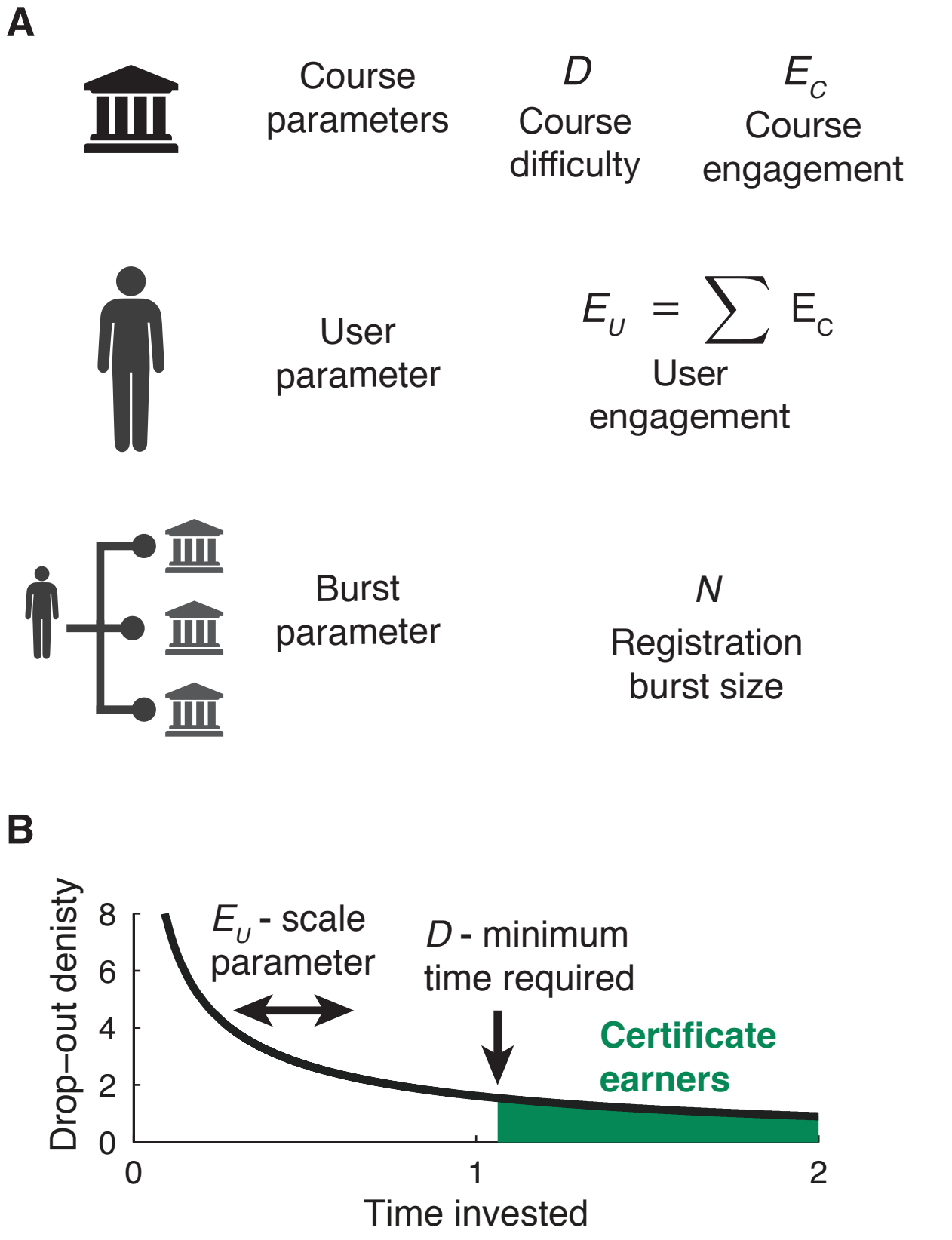

**Figure 1 The *βEND* model parameterizes a Weibull survival function and accounts for the effect of multiple registrations. A)** Each registration event is associated with a course difficulty, *D*, and a user engagement $E_U$. User engagement is estimated as the sum of course engagement, $E_C$, for all courses registered by a user during the study period. Users may register for multiple courses in a single burst event, with *N* the number of registrations. **B)** The $E_U$ and *D* terms have intuitive interpretations in a Weibull probability density function. Each student who registers for a course remains enrolled for a certain time before dropping out. The *D* term can be considered a minimum time investment with which a student can earn a certificate. The *D* term is therefore analogous to course difficulty. The $E_U$ term functions as a scale parameter that shifts the distribution toward higher investment times. We therefore interpret $E_U$ as a representation of user engagement.

### Model parameters were fit by maximum likelihood

The *βEND* model seeks to describe the effect of multiple registrations on certificate rates. However, in the FUN dataset, 56% of users register for only one course. For these users, $E_U = E_C$ and equation 1 rearranges to:

$$E_C = \frac{D}{-log(S)^{\frac{1}{\beta}}} \quad (2)$$

We used this relation to constrain $E_C$ as a function of *D* for each course. Because our model incorporates the observed baseline certificate rate for single-registered users of each course, the predictive power is limited the changes in certificate rate expected from multiple registrations.

In total we fit 92 free parameters: 91 values of $D$ for the certificate-offering courses in our dataset plus a single value of $\beta$, the Weibull shape parameter. Cross-validation was performed by randomly assigning users to training or test groups. The model was fit to the certificate status of ~330,000 training registration events of users by maximum likelihood. All reported statistics of the performance of the model are on users of the test group.

## Registration events can be clustered into well-defined bursts

To determine values for $N$, the registration burst size, registration events for each user were clustered in time. The histogram of same-user registration delay times revealed a bimodal distribution (Supplemental Fig. S2A). Re-registration rates reached a local minimum after a delay of around 8 hours, with most re-registrations occurring after either significantly shorter or longer delay times. We therefore set a delay threshold of 8 hours to define registration burst events by agglomerative clustering. Varying the clustering threshold between 4-24 hours changed the total number of identified clusters by less than 1% (Supplemental Fig. S2B).

## Co-registration affects expected certificate rates

Figure 2 compares certificate rates derived from the *βEND* model with rates in the FUN dataset. Registration events were assigned certificate probabilities over two orders of magnitude that agreed well with empirically observed rates. In contrast, a conventional logistic regression model systematically overestimated certificate probabilities for low-probability events (Fig. 2A).

We define a co-registration cohort as an ordered pair of two courses and the set of users who have registered for both. Users may also have registered for additional courses. The conditional certification rate for the cohort is the certificate rate for the first course among users who also registered for the second. Of 8190 possible co-registration cohorts, 7321 produced at least 5 certificates and were included in the analysis.

The certificate rates of each co-registration cohort were well described by the *βEND* model (Fig. 2BC). Predicted certificate log-odds correlated with observed values (Pearson's $r^2$ = 0.72). We also examined the change in certificate rates for each course associated with co-registration in each other course, expressed as a log odds ratio (Fig 2C). Empirically observed log odds ratios correlated with model-derived values ($r^2$ = 0.59).

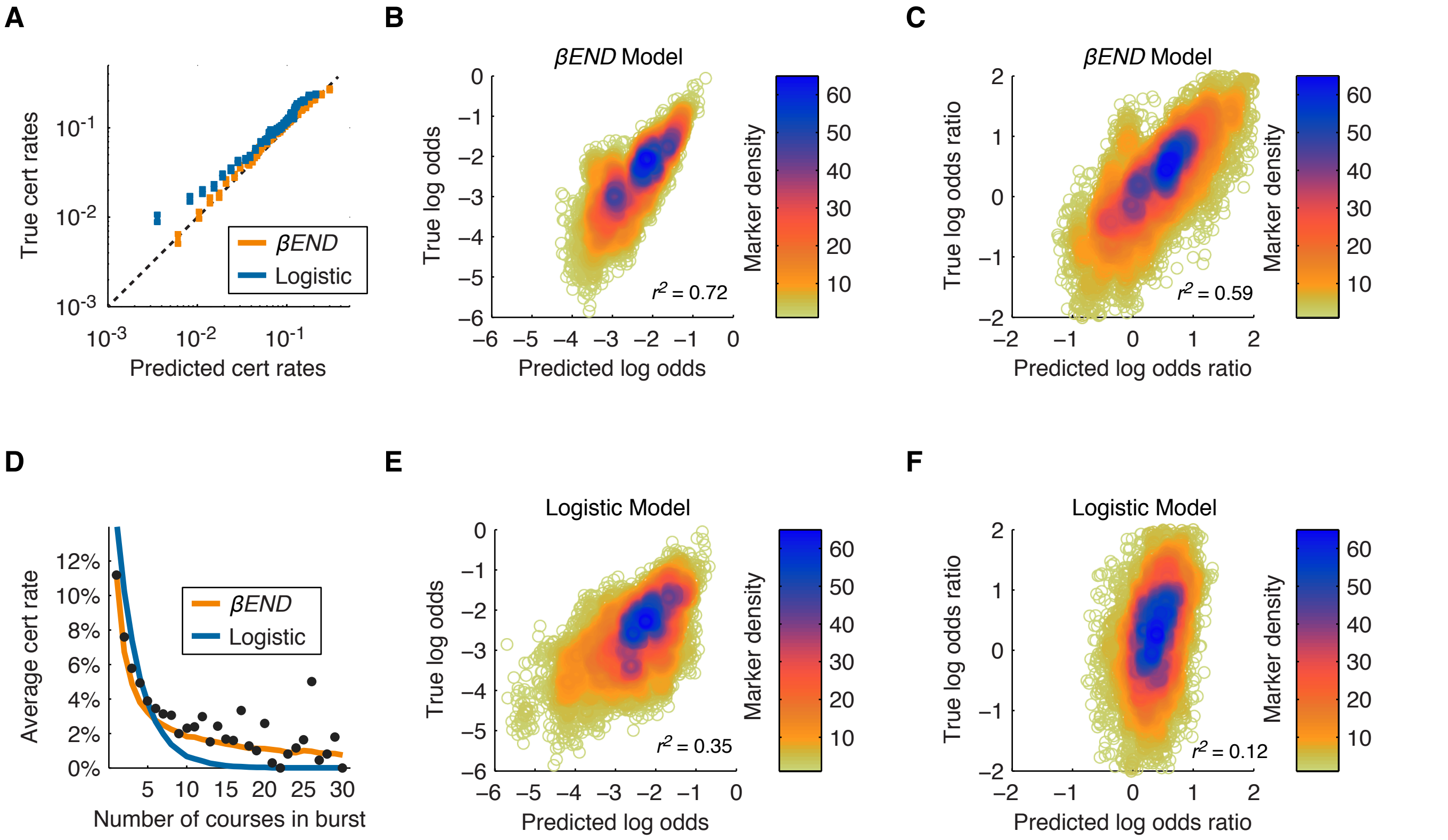

**Figure 2 The *βEND* model outperforms logistic regression in predicting key features of the FUN dataset. A)** A Hosmer-Lemeshow plot comparing certificate rate predictions of the *βEND* and logistic models. All recorded registration events were sorted into 20 bins by predicted certificate rate. The mean predicted and true observed certificate rate for each bin is plotted for both models. While the *βEND* model predictions show a linear relation over two orders of magnitude, the logistic model systematically overestimates certificate rates for low-rate registration events. **B)** Users were grouped by co-registration cohort, defined as the set of users who registered for each possible set of two courses. The *βEND* model was used predict the log odds of obtaining a certificate in the first course conditioned on co-registration in the second. Predicted and observed log odds values were linearly correlated ($r^2$ = 0.72). **C)** Log odds ratios for each co-registration cohort were calculated as the log odds of obtaining a certificate given co-registration minus the log certificate odds for users who registered for only the course alone. Log odds ratios predicted by the *βEND* model correlated with observed values ($r^2$ = 0.59). **D)** Registration events were grouped by burst size, the number of registrations recorded by a user within a short time period. The approximately inverse relationship between certificate rate and burst size was well described by the *BEND* model but not by conventional logistic regression. **E)** Predictions obtained through logistic regression for the certificate log odds of co-registration cohorts were less well correlated to observed log odds ($r^2$ = 0.35). **F)** Similar predictions obtained by logistic regression for certificate log odds ratios relative to single-registered users correlated with observed values ($r^2$ = 0.12).

Figure 2D shows certificate rates as a function of registration burst size. The per-course certificate rate drops in inverse proportion to the number of simultaneously co-registered courses. The effects of burst size on certificate rate were captured by the $\beta END$ model but overestimated by logistic regression. The logistic regression model also performed relatively poorly in predicting the certificate log-odds (Fig 2EF).

### Engagement governs the effect of co-registration

Equation 1 can be log transformed twice to isolate the respective contributions of difficulty and engagement to the certificate rate.

$$log(-log(S)) = \beta \cdot log(D) - \beta \cdot log(E) \quad (3)$$

We made use of this linearization to express the effect of co-registration as the loglog transformed certificate rate of each co-registration cohort minus the loglog transformed rate of users who registered for only a single course. Because both sets of registration events share the same course difficulty, the result reflects the difference in user engagement between the single-registered and the co-registered populations.

Figure 3A compares the certificate rate of users who registered for only a single course with that of every possible co-registration cohort. 32% of courses saw significant increases in certificate rate given co-registration while 35% saw decreases. These changes were homogenous across the set of possible co-registrations. In other words, courses that benefited from co-registration tended to benefit from co-registration with any other course.

The global effect of co-registration on double-log transformed certificate rates was largely governed by log user engagement (Fig 3B). In courses with below average user engagement, co-registration will generally increase expected engagement levels. In high-engagement courses, co-registration will generally decrease expected engagement and therefore certificate rates.

### Engagement levels vary with time and age

The $\beta END$ model assigns user engagement and course difficulty scores to each registration event. In this way, it allows user-specific and course-specific factors affecting user registration to be decoupled and separately related to other relevant data.

We first looked at the changes in model-derived parameters for users who returned to the platform multiple times during the data collection period (Fig. 4ACE). Two or more independent registration burst events were recorded for 30% of users. Certificate rates did not change significantly following re-registration (Fig. 4A). However, both course difficulty and course-associated engagement levels were found to significantly increase (Fig. 4CE).

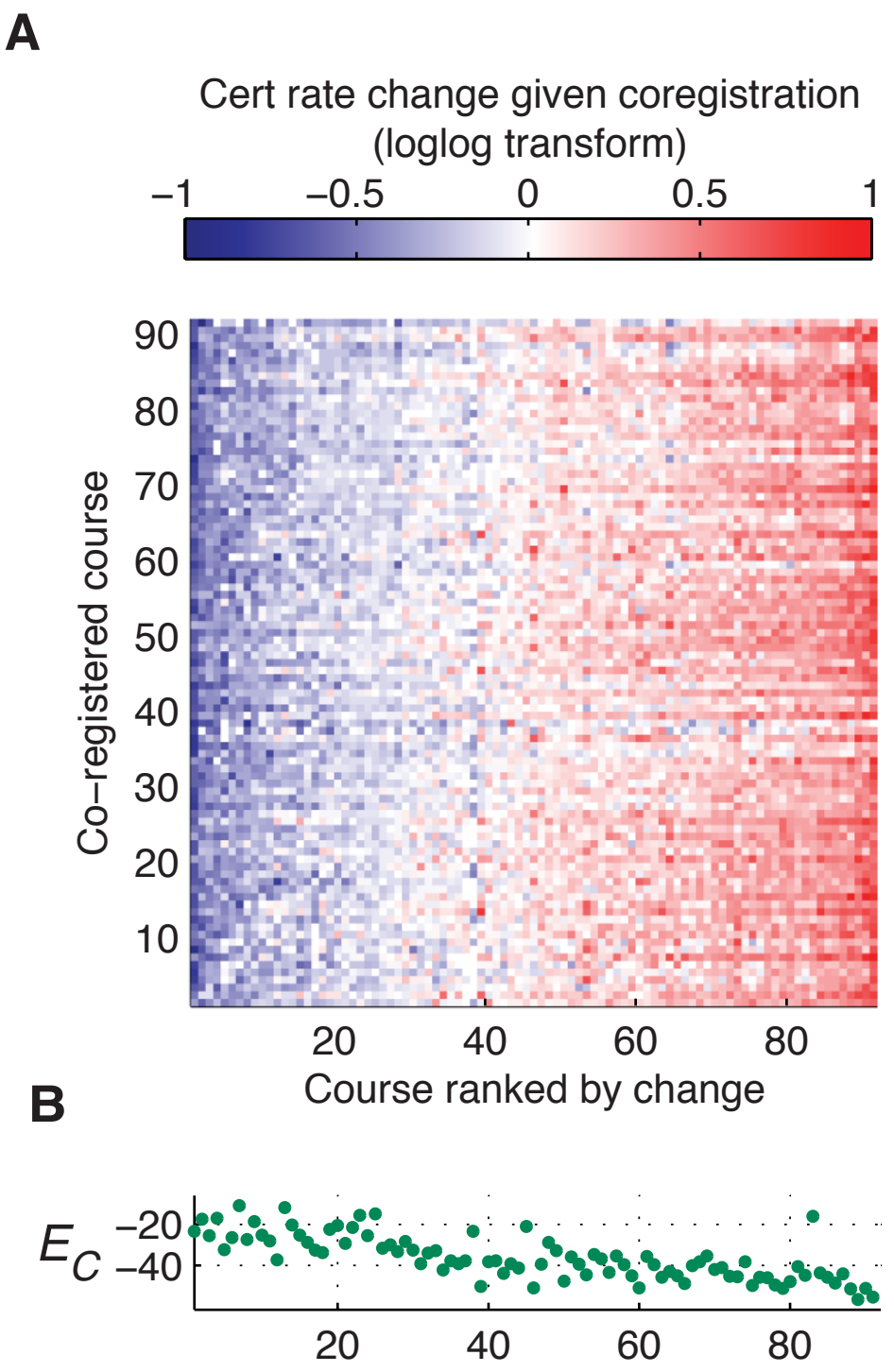

**Figure 3 Course-course interactions are uniform with respect to the co-registered course and are governed by user engagement.** **A)** Heatmap showing the effect of co-registration on certificate rates. The effect is expressed as the loglog transformed certificate rate of a given course, conditional on co-registration with each other course, minus the loglog transformed certificate rate of users who registered for only each course alone. The registered courses indicated on the horizontal axis were ordered by the median effect of co-registration. **B)** Model-derived course engagement values for each course, with courses ordered as above. Courses with low engagement courses benefit the most from co-registration, regardless of which other course was co-registered.

The FUN dataset also includes a self-reported age for 93% of users. Certificate rates increase consistently with user age (Fig. 4B). This increase is associated with a decline in course difficulty between the ages of 20 and 50 (Fig. 4D). Following age 55 we observed a significant increase in the engagement term (Fig. 4F). Therefore, the positive trend in certificate rates is driven first by course difficulty, then by user engagement at later ages.

## DISCUSSION

Recent debates have focused on the causes of putatively low MOOC certificate rates(*13*). We found that a significant drop in average certificate rate is linked to bursting registration behaviors. Bursting registrations are likely to be common on other MOOC platforms because activity bursts are a general feature of internet user behavior(*14*).

This result suggests that certificate rates could be increased by constraining burst sizes, for example by limiting users to one new registration per day. However, it is not clear that users would be well served by this. Registration bursts might represent a kind of course-shopping strategy through which users optimize their course selections(*15*). As evidence of this, we observed that users who register in larger bursts earn more total certificates, even as their average certificate rate declines.

If many burst registrations are never invested with serious user effort, then the average certificate rate will not reflect a typical user's experience. Instead, we propose certificates-per-burst as a simple metric for the expected overall success of each user upon each approach to the platform. The per-burst certificate rate for FUN is 12.2%, 1.5-fold higher than the raw certificate rate.

The *βEND* model outperformed logistic regression in fitting this dataset. The symmetrical logit link function is known to exhibit bias when applied to asymmetrical datasets where the chance of success approaches zero much faster than one(*16*). Previous work has shown generalized Weibull linkage functions to outperform logistic approaches in describing mechanical failure, mortality, and other systems characterized by weakest-link scaling(*17*).

The success of the *βEND* model demonstrates the information richness of complete user registration profiles. However, similar datasets may be difficult to obtain for other user populations. While FUN is the predominant French-language MOOC platform, several platforms compete for the attention of English-speaking students. Even within a single platform, user data may be segregated among participating institutions, constraining analyses to the institution level(*18*). Any comprehensive MOOC performance model should thoroughly account for co-registration, which may occur across platforms.

The best-fit value for Weibull shape parameter $\beta$ was 0.13, indicating that the drop-out rate does not follow the dynamics of conventional aging. Instead, values of $\beta < 1$ indicate that survivors become more reliable over time. The MOOC user population could be said to experience burn-in, with users who persist in a course becoming increasingly inclined to finish it. Other work has shown that investments of user time and effort produce non-linear increases in certificate rate(*5, 7, 18*).

We found co-registration was associated with significant changes in certificate rate. Out of 91 courses, 27 saw a 2 fold increase or more in certificate rate for users that had co-registered another course. Another 31 courses saw a 2 fold decrease in rates given co-registration. Remarkably, the certificate rate changes were broadly similar regardless of which other course was co-registered.

The effects of co-registration were mediated by course engagement levels. Our model associates each course with an engagement score and estimates a user's engagement as the average of their registered courses'. In courses with below average engagement, co-registration tends to raise a user's expected engagement level. The opposite is true for courses with above average user engagement, where co-registration is associated with lower certificate rates.

Users who returned to the FUN platform to take new courses were not more likely to earn certificates, even after multiple rounds of re-registration. This is surprising given the expectation that users become more knowledgeable and skilled through coursework. The model-guided analysis revealed that returning users select courses of progressively higher engagement scores and higher difficulty, two opposing trends that result in no net change in certificate rate.

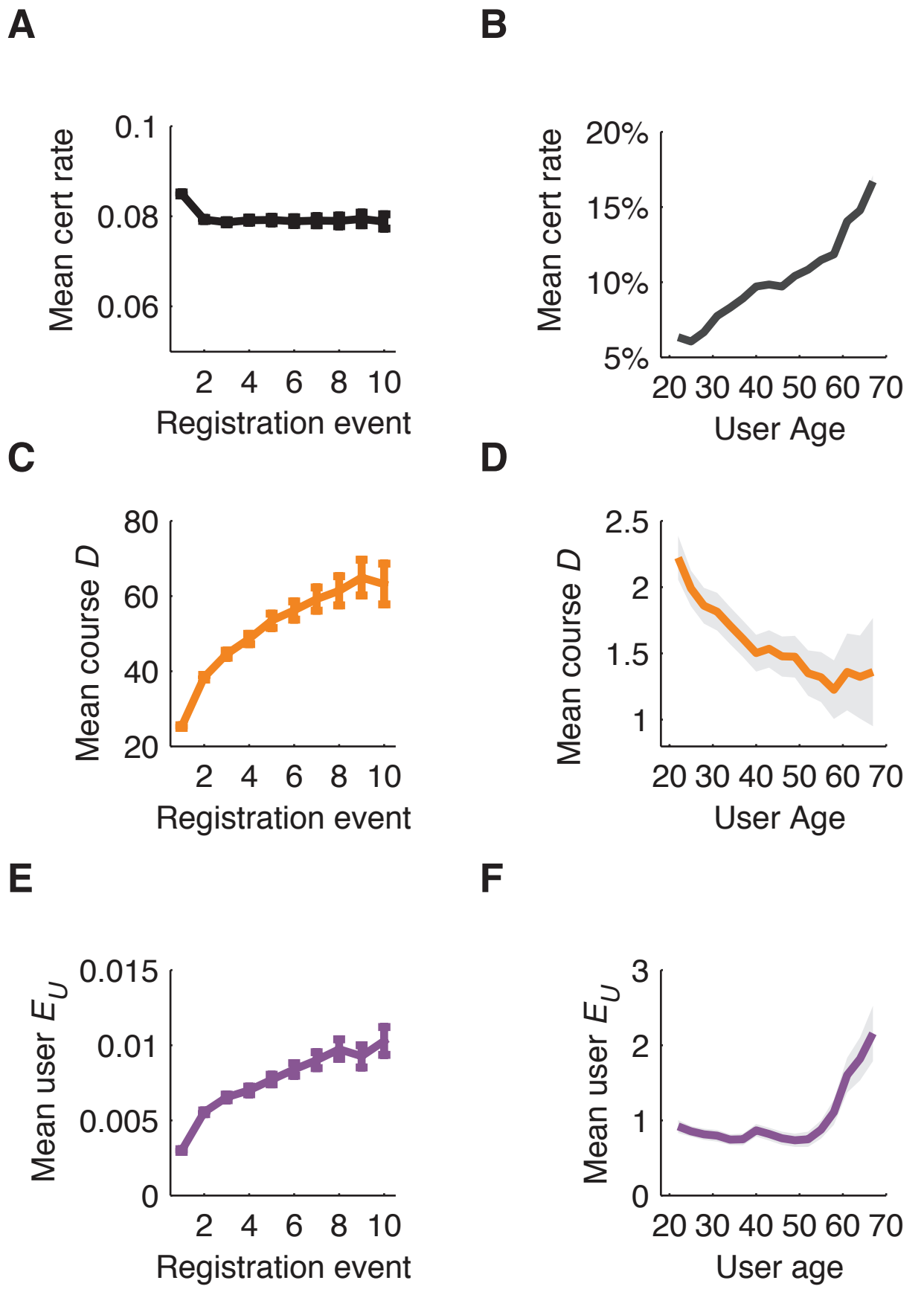

**Figure 4 Model-derived difficulty and engagement values provide mechanistic insight into trends in user certificate rates. ACE)** Users who return to the platform for multiple registration bursts do not show significant increases in certificate rate. The *βEND* model indicates that both user engagement and course difficulty consistently increase with re-registration. Increases in user engagement scores may reflect positive learning outcomes. **BDF)** Certificate rates increase significantly and monotonically with user age. The *βEND* model suggests this pattern is driven by decreasing course difficulty from younger to middle aged users, then by increasing engagement among older users.

This result underscores that certificate rates alone should not be used to track user progress because they also reflect the difficulty of the selected courses, which may systematically vary between groups and over time. The engagement term of the *βEND* model effectively controls for course difficulty and reports only user-dependent contributions to success. In the case of returning users, increasing engagement could reflect progressive learning acquired through coursework. Alternately, it may indicate a tendency of users who are a priori more engaged to concentrate in specific courses as they become familiar with the platform.

Decoupling course difficulty from user engagement provides insight into other demographic trends associated with user success. Overall certificate rates increase monotonically with user age. Between youth and middle age, the trend is driven by decreasing course difficulty. This could mean that middle-aged users value certificates more highly and select courses where they are more likely to obtain one. Alternately, it could indicate that courses appealing to middle-aged professionals are less technical or challenging than courses selected by college-aged MOOC users. The ages of 45-65 are marked by a progressive increase in user engagement. This might imply more time, greater ability, or simply more enthusiasm for digital learning among older users.

A complete understanding of the factors that promote effective web-based learning will require major new programs in education research(*19*). Many of the concepts currently being used to describe digital experiences were developed for the pre-digital era of distance education(*20-22*). MOOC datasets are not only much larger, but document new kinds of behaviors and relationships that cannot be easily described with established theories(*23*). New theoretical frameworks are needed to describe the fluid, networked experience of taking an online course online and generate actionable models for making those courses better.

The $\beta END$ model formally connects MOOC certificate rates to systems reliability theory. Reliability engineering deploys model-guided interventions to reduce failure rates in complex systems. Future efforts to improve MOOC user outcomes may benefit from the quantitative framework applied to other systems characterized by stress, fatigue and random break-downs.

## ACKNOWLEDGEMENTS

We thank Catherine Mongenet and the FUN platform for providing access to the anonymized MOOC registration data. Dusan Misevic and Ignacio Atal provided feedback on the structure and content of this manuscript. Marc Santolini and Liubov Tupikina provided advice on analyzing and visualizing co-registration data. **Author Contributions:** EHW, MC and ABL planned the study and analyzed the results. MC obtained and curated the dataset. EHW derived the $\beta END$ model. EHW, MC and ABL wrote the manuscript. **Funding:** This work was supported by a CRI Research Fellowship to EHW. **Competing Interests:** The authors declare no competing interests.

SUPPLEMENTARY INFORMATION FOR

# A survival model for course-course interactions in a Massive Open Online Course platform

Edwin H. Wintermute, Matthieu Cisel & Ariel B. Lindner

Corresponding Author: Edwin H. Wintermute

Email: ehwintermute@gmail.com

THIS PDF FILE INCLUDES

- Supplementary text
- Figs. S1 to S6
- References for SI reference citations

## SUPPLEMENTARY METHODS

### Data Collection and Anonymization

Course registration and user certificate profiles were collected from October 2013 to September 2015 using the Google Analytics platform. Registrants were asked to self-report their gender, birth year, country of residence and highest level of education attained. Personally identifying information including usernames, full names and email addresses were removed prior to analysis. The privacy policy and terms and conditions of use for the France Université Numérique are available on the platform website (www.fun-mooc.fr).

### Definition and Characterization of Burst Registration Events

Figure S1 depicts registration events over time for representative users. Registrations were not uniformly distributed across the study period, but tended to cluster in time. We sought to quantify this pattern and assign individual user registrations to well defined registration burst events. Figure S2 shows the distribution of delay times between same-user registration events. The large majority of re-registrations occurred within a few hours of a previous registration event. A significant number users also returned to the platform after a delay of weeks or months. However, re-registration delays of intermediate periods, from 2-24 hours, were relatively rare. The bi-modal nature of the delay distribution suggests that most registration events can be assigned to short-term bursts of activity that are separated by relatively long periods of inactivity.

Registration burst events were defined using agglomerative clustering in MATLAB. A hierarchical cluster tree was generated using the time between registration events as a distance metric and a nearest-neighbor linkage function. Bursts were separated from the cluster tree using a delay time cutoff of 6 hours. Varying the chosen cutoff time from 2-24 hours had little effect (<1%) on the total number of bursts defined (Fig. S3).

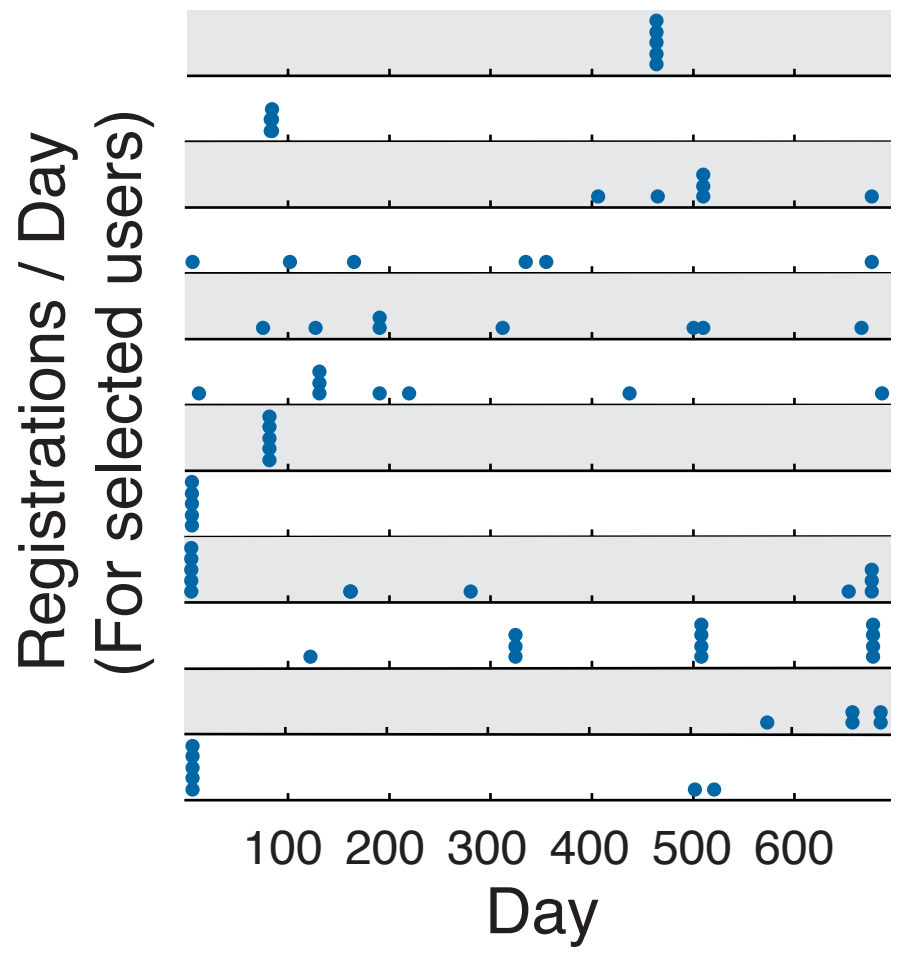

**Figure S1 User registrations often come in bursts.** Each row depicts the registration activity of a representative user. Blue dots indicate registration events. Dots were stacked vertically to indicate multiple registrations on the same day.

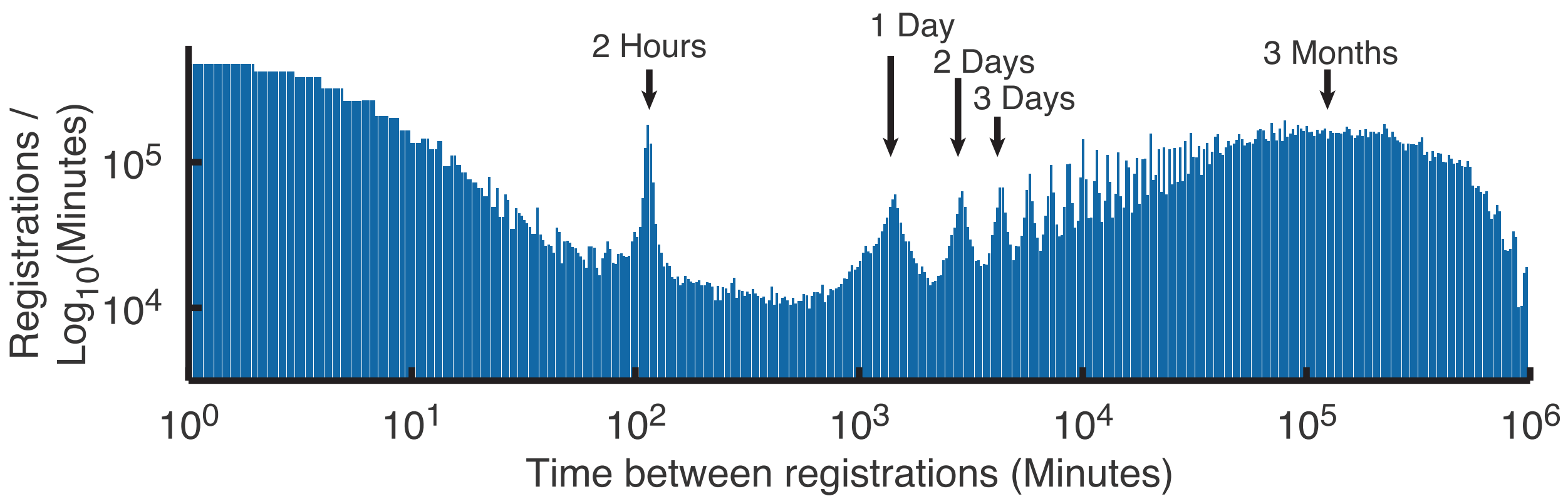

**Figure S2 Registration delay times follow a roughly bimodal distribution.** Most re-registrations occurred either within a few hours of a previous registration event or after a delay of several weeks. Arrows indicate time points of interest. The day/night cycle of user activity can be seen in the peaks recurring at intervals of 24 hours.

The assignment of registrations to burst events created two new metrics. Each user is associated with a burst number, defined as the total number of burst events recorded for that user during the study period. Each burst is associated with a burst size, the total number of courses registered during the event. We characterized the distribution of these quantities among users of the FUN platform.

Both burst size and burst number were found to follow heavy-tailed distributions that were well described by a modified power law (Fig. S4). Power law distributions are commonly found in internet user activity (*1*). In this case, deviation from a strict power law may be explained by the fact that the total number of available courses, and therefore the maximum number of registrations, is bounded at a maximum of 140.

The heavy-tailed character of these distributions indicates that a relatively small number of super-users are responsible for an outsized portion of registration activity. For example, 70% of FUN users were associated with only a single registration burst event. Only 15% of users were associated with 3 or more registration bursts, yet these users were responsible for 63% of the total recorded bursts. Similarly, only 12% of registration bursts were of size 3 or more, yet these bursts accounted for 35% of total recorded registrations.

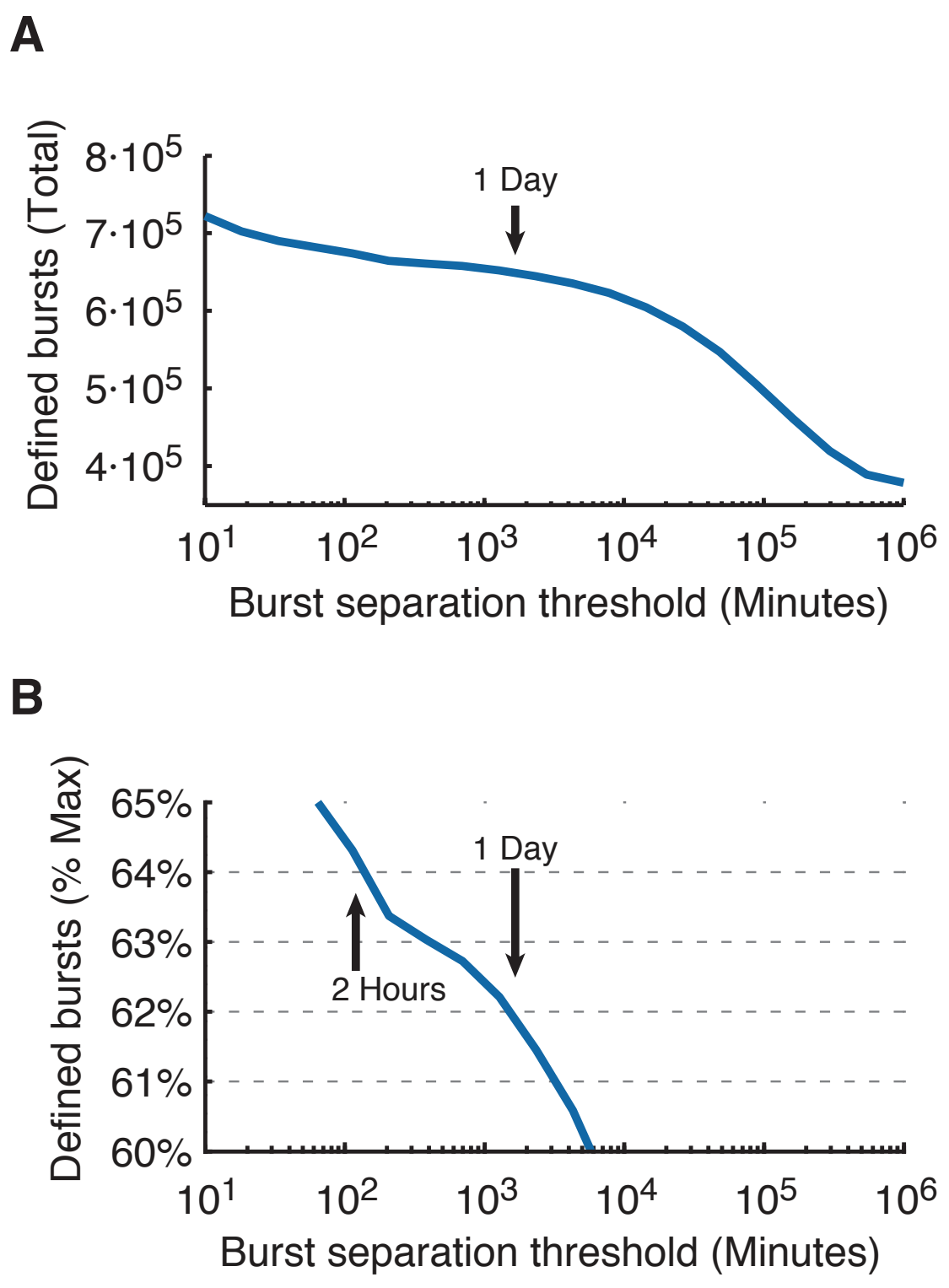

**Figure S3 Definition of registration burst events is insensitive to the selected clustering threshold.** User registrations were assigned to registration burst events using an agglomerative clustering algorithm with a varying burst separation threshold. For each threshold, the total number of defined burst events is indicated. **A)** The total burst count changes slowly with respect to the burst separation threshold in the vicinity of one day. **B)** Varying the burst separation threshold between 2 hours and 1 day changed the total number of defined burst events by less than 1%. This indicates that the assignment of registrations to burst events is robust to the chosen clustering threshold. A clustering threshold of 6 hours was selected for further analysis.

The $\beta END$ model weights each registration event equally, regardless of the total registration activity of the associated user. Super-users may therefore have a significant influence on the global model behavior.

### Impact of Burst Registration Dynamics on Certificate Rates

The $\beta END$ model characterizes registration bursts with a single parameter, $N$, the burst size. Our dataset also records the chronological order of burst events and the order of registrations within each burst event, but these parameters were not used to estimate certificate rates. We examined the relationship between registration order and certificate rate to test the assumption of order independence.

Figure S5 depicts the average course certificate rate for burst events, and registration events within a burst, ordered by their occurrence in time. Certificate rates were higher for a user's first registration burst but did not change further over second and later bursts. Similarly, the first course registered within a burst is more likely to be certified than subsequent courses. First registrations may be associated with higher user interest or motivation because they were the original inspiration for the registration activity.

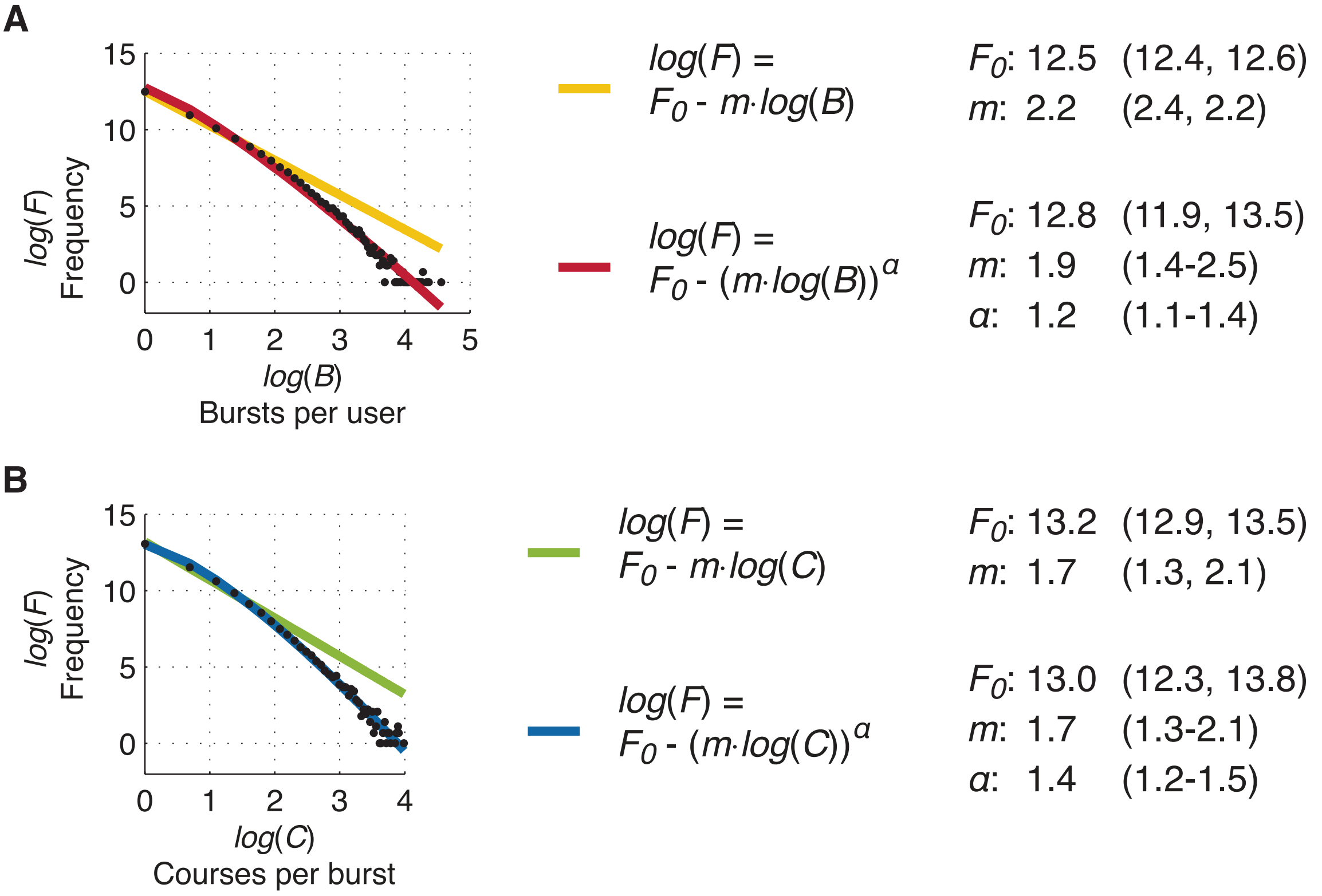

**Figure S4 Burst metrics follow a modified power law. A)** Frequency distributions and model parameters for the burst number, defined as the total number of registration burst events recorded for each user. **B)** Frequency distributions and model parameters for the burst size, defined as the number of course registrations recorded for each burst event. Neither distribution was well described by a strict power law but both could be fit with an additional term producing a curved power law. Best-fit parameter values are indicated for each model as well as 95% confidence intervals.

However, the impact of registration order was small relative to the effect of burst size, supporting their omission from the simplified *βEND* model. Future models may incorporate more details of user behavior to produce more precise certificate rate estimates.

We next sought to characterize the influence of burst size and burst number on user certificate rates (Fig. S6). Users who registered for more courses had slightly higher certificate rates on average, an effect which quickly leveled off. Increasing user registrations may take the form of additional bursts, larger bursts, or both. Larger burst sizes were associated with reduced certificate rates. Conversely, certificate rates increased with user burst number.

These results are consistent with a model in which users divide their attention among multiple courses registered simultaneously, resulting in a lower average certificate rate. On the other hand, users who return to the platform following a significant delay may have a greater interest in or motivation for online learning. Taken together, these trends indicate that the highest certificate rates are found in users who register for many, small, bursts. The opposite effects of burst size and burst number underscore the importance of separating these metrics when seeking to predict user certificate rates.

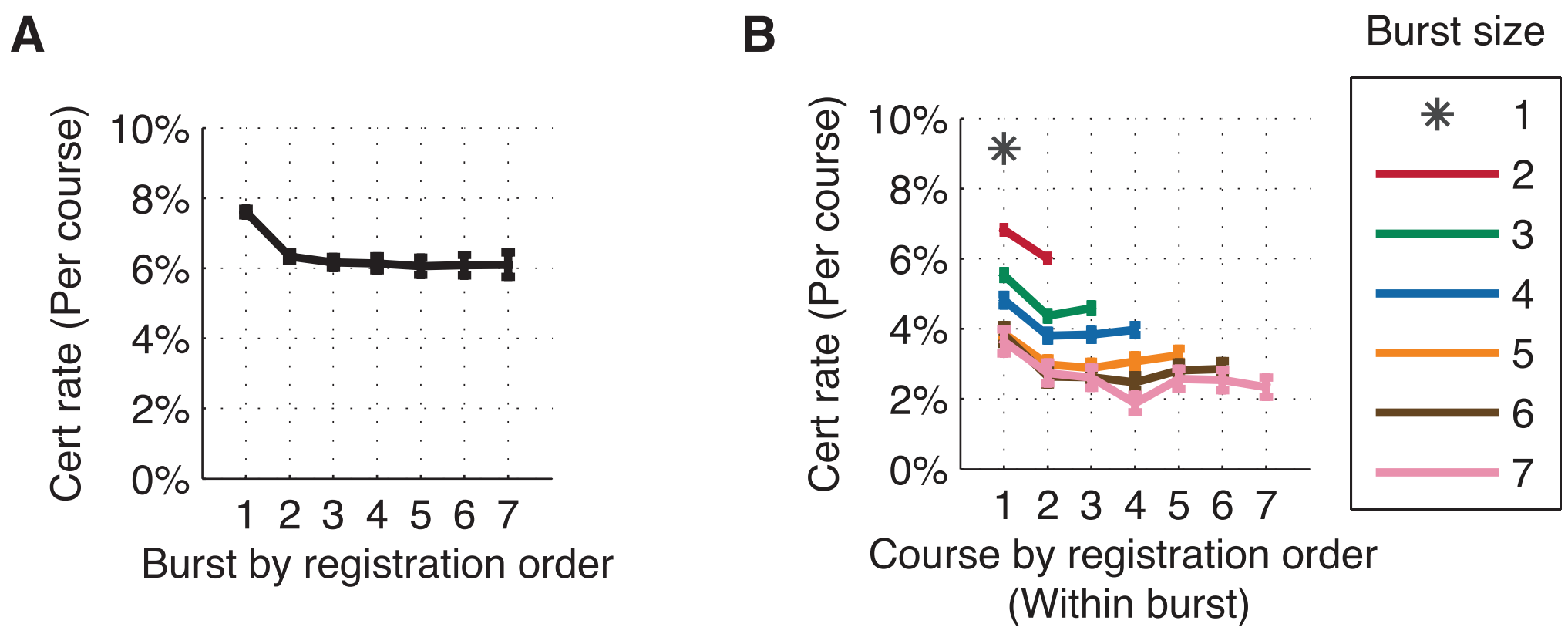

**Figure S5 Certificate rates vary little with registration order. A)** Average per-course certificate rates are shown as a function of burst registration order, determined chronologically. Certificate rates were significantly higher for a user's first registration burst event, then plateaued for additional events. **B)** Average certificate rate is depicted as a function of chronological registration order of individual courses within bursts of varying size. Error bars indicate 95% confidence intervals.

### Detailed Derivation of the *βEND* Model

The personal investment of a user in a course is expressed in units of pseudo-time, $\hat{t}$, which may be considered as the true quantity of time spent with an unknown scaling factor to reflect personal productivity. A user may withdraw from a course after investing any amount of pseudo-time, following a probability density function $P(\hat{t})$. User withdrawal is a passive event and not associated with any recorded action on the MOOC platform.

Following reliability models of aging, we define a survival function, $S(\hat{t})$, as the probability that the withdrawal time for a given user, $\hat{T}$, is greater than $\hat{t}$. The survival function is simply $1 - F(\hat{t})$ where $F(\hat{t})$ is the cumulative distribution function of $P(\hat{t})$.

$$S(\hat{t}) = P(\hat{T} > \hat{t}) = 1 - P(\hat{T} < \hat{t}) = 1 - F(\hat{t}) \quad (1)$$

The hazard rate, $H(\hat{t})$, is the instantaneous relative risk that a student will withdraw from a course. It represents the withdrawal rate at $\hat{t}$ conditioned on a user having persisted for at least $\hat{t}$ units of time in the course. The hazard rate is obtained from the survival function as follows.

$$H(\hat{t}) = \frac{dS}{d\hat{t}} \cdot \frac{1}{S} = \frac{-d\log(S)}{d\hat{t}} \quad (2)$$

In the Weibull model for aging systems, the hazard rate changes with time as a power function.

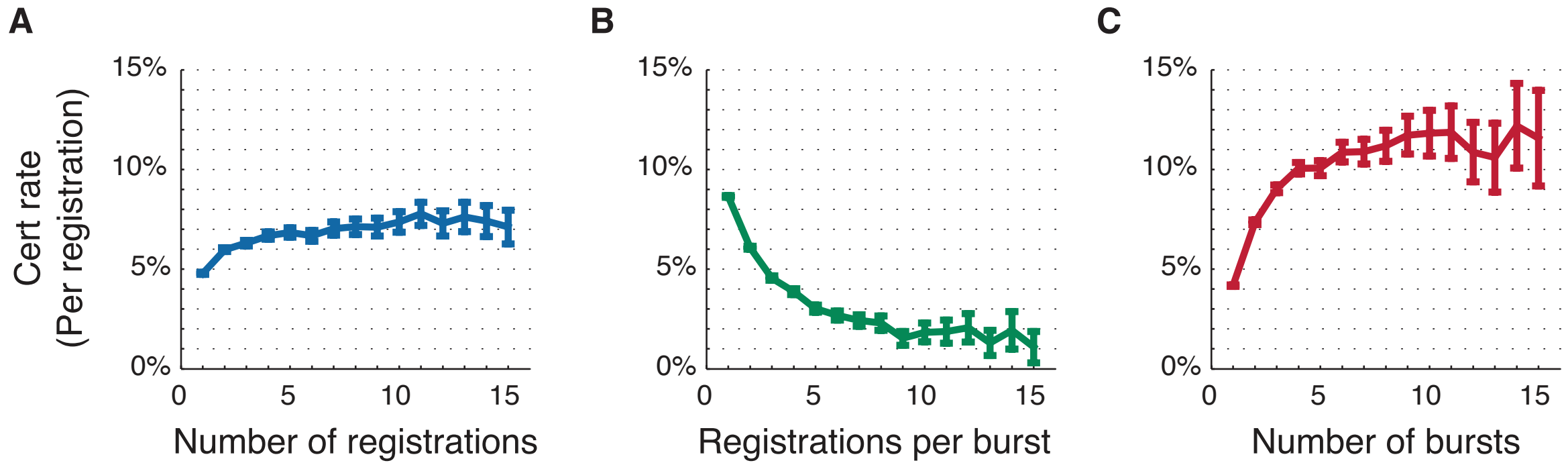

**Figure S6 Burst size and burst number have opposing effects on certificate rate. A)** A user's average certificate rate increases gradually with their total number of platform registrations. The effect levels off after roughly five total registrations. **B)** Certificate rates decrease with burst size. **C)** Certificate rates increase with burst number. Error bars represent 95% confidence intervals.

$$H_W(\hat{t}) = \frac{\beta}{E^\beta}\hat{t}^{(\beta-1)} \tag{3}$$

The term $\beta$ is the shape parameter for the Weibull distribution. The values of $E$, $\hat{t}$, and $\beta$ are all constrained to be positive. Note that for $\beta = 1$ the hazard rate is constant in time and the Weibull model reduces to an exponential model. For $\beta > 1$ the hazard rate increases with time and the system is said to experience aging. For $\beta < 1$ the hazard rate slows with time. In our context, this means that the chance that a user will withdraw from a course decreases as the user invests more time.

From the Weibull hazard rate $H_W$ we obtain the Weibull survivor function $S_W$

$$S_W(\hat{t}) = exp\left[-\int_0^{\hat{t}} H_W(x)dx\right] = exp\left[-\left(\frac{\hat{t}}{E}\right)^\beta\right] \tag{4}$$

The difficulty of a course, $D$, is defined as the minimum time investment required to obtain a certificate. Users who disengage from a course at any time prior to $D$ will not certify it. The probability of obtaining a certificate is therefore calculated as the survival function taken at $\hat{t} = D$.

$$P(\hat{T} > D) = exp\left[-\left(\frac{D}{E}\right)^\beta\right] \tag{5}$$

Finally we account for the fact that a user may have registered for N courses simultaneously during a registration burst. In this case, we simply assume that a user's certificate probability will decline in inverse proportion to $N$. This is represented as a $-log(N)$ term within the exponential expression for $S$.

| | |
|---|---|
| $\hat{t}$ | The pseudo-time invested by a user in a course. |
| $\hat{T}$ | The pseudo-time at which a given user withdraws from a course. |
| $\beta$ | A Weibull shape parameter. |
| $E$ | The engagement level of a user. |
| $N$ | The number of courses registered in a single burst. |
| $D$ | The difficulty of a course; the minimum pseudo-time investment required to certify. |
| $P(\hat{t})$ | A probability density function. |
| $F(\hat{t})$ | A cumulative distribution function. |
| $P(\hat{t})$ | A survival function. |
| $S_W(\hat{t})$ | A Weibull survival function. |
| $H(\hat{t})$ | A hazard rate function. |
| $H_W(\hat{t})$ | A Weibull hazard rate function. |

**Table 1. Summary of notation used in construction of the *βEND* model.**

$$S = exp\left[-\left(\frac{D}{E_U}\right)^{\beta} - log(N)\right] \quad (6)$$

**Parameter Fitting for the *βEND* Model**

We calculate $C_j^i$, the probability that user i obtains a certificate for course j following a certain registration event. This calculation requires values for $E_U^i$, the engagement level of user i; $D_j$ the difficulty of course j; and $N_k^i$, the number of courses registered by user i during k, the registration burst that includes course j.

The difficulty score $D_j$ is fit as a free parameter for each course. We also fit $\beta$ as a global Weibull shape parameter. Thus our model contains 92 total free parameters: 91 values of $D$ and a single $\beta$.

From $D_j$ and $\beta$, we calculate $E_C^j$, the engagement score for course j. This value is constrained by equation 6. Here we make use the singleton certificate rate, $C_S^j$, which is defined as the certificate rate for users who registered for course j and no other courses. This allows us to take $N = 1$ and arrive at the following relation.

$$E_C^j = \frac{D_j}{-log(C_S^j)^{\frac{1}{\beta}}} \quad (7)$$

| | |
|---|---|
| $C_j^i$ | The probability that registration of user i to course j produces a certificate. |
| $E_U^i$ | The engagement level of user i. |
| $E_C^j$ | The engagement coefficient of course j. |
| $D_j$ | The difficulty level of course j. |
| $\{R^i\}$ | The set of courses registered by user i. |
| $\{Z_k^i\}$ | The set of courses registered by user i during registration burst k. |
| $N_k^i$ | The total number of courses registered by user i during registration burst k |
| $C_S^j$ | The certificate rate for singleton users who registered only course j. |

**Table 2. Summary of notation used for parameter fitting.**

Each user, i, is associated with a user engagement score, $E_U^i$. User engagement scores are estimated as the sum of the course engagement scores over $\{R^i\}$, the set of all courses registered by user i in the dataset.

$$E_U^i = \sum E_C^j \qquad j \in \{R^i\} \tag{8}$$

A burst event is one subset of registration events for a particular user that are clustered in time. We collect $\{Z_k^i\}$, the set of courses registered by user i during their kth registration burst. $N_k^i$ is the total number of courses registered in this event.

$$Z_k^i \subseteq \{R^i\} \quad , \quad N_k^i = \left|\{Z_k^i\}\right| \tag{9}$$

Finally, we obtain the certificate probability $C_j^i$, for the registration to course i by user j.

$$C_j^i = exp\left[-\left(\frac{D_j}{E_U^i}\right)^{\beta} - log\left(N_k^i\right)\right] \tag{10}$$

Optimal values for $D$ and $\beta$ were obtained with the method of maximum likelihood. The parameter search was conducted using the derivative-free simplex method implemented as the fminsearch function in MATLAB (*2*).

### Derivation and Parameter Optimization for the Logistic Model

A logistic model was constructed as a control and benchmark for the performance of the $\beta END$ model. Following standard practices, the log-odds of course certification, $L$, were estimated as a linear combination of course difficulty, $D_C$ user engagement, $E_U$, and burst size, $N$.

$$L = D_C - E_U - \gamma N \tag{11}$$

As with the $\beta END$ model, we estimate user engagement as the sum of course engagement terms, $E_C$, for the set of courses registered by the user during the study period, $\{R^i\}$.

$$E_U^i = \sum E_C^j \qquad j \in \{R^i\} \tag{12}$$

We then made use of the set of single-registered users to derive $E_C$ as a function of $D_C$ reducing the dimensionality of the model.

$$E_C^j = D_C^j + \gamma + L_S^j \tag{13}$$

Where $L_S^j$ is the log certificate odds of singleton users who registered for course j and no other courses.

Optimal values for $D_C$ and $\gamma$ were generated with the method of maximum likelihood using the fminsearch function of MATLAB. Both the $\beta END$ model and the logistic model were fit with 92 free parameters.

## SUPPLEMENTARY REFERENCES